\documentclass[preprint,12pt]{elsarticle}

\usepackage[a4paper, total={6in, 9in}]{geometry}
\usepackage{graphicx}  
\usepackage{dcolumn}   
\usepackage{bm}        
\usepackage{amssymb}   
\usepackage{amsmath}
\usepackage{braket}
\usepackage{enumitem}
\usepackage[mathscr]{euscript}
\usepackage{color}
\usepackage{epsfig,color}
\usepackage{hyperref}
\DeclareUnicodeCharacter{3B1}{\ensuremath{\alpha}}

\journal{Chaos, Solitons and Fractals}

\begin{document}
	
	\begin{frontmatter}
		
		\title{Anti-Phase Synchronization of Chaos in $\mathcal{PT}$-Symmetric Nonlinear Oscillators}
		
		\author[inst1]{Jyoti Prasad Deka}
		\ead{jyoti.deka@alumni.iitg.ac.in}
		
		\affiliation[inst1]{organization={Department of Physics},
			addressline={Girijananda Chowdhury University, Azara}, 
			city={Guwahati - 781017},
			state={Assam},
			country={India}}

		\begin{abstract}
			We investigate the temporal dynamics of the $\mathcal{PT}$-Symmetric nonlinear oscillators in the presence of Duffing nonlinearity for two forms of oscillator configuration. In the former, we consider two oscillator coupled to each other. One oscillator is amplified and the other is attenuated. From the bifurcation analysis, we find that the temporal evolution of oscillators exhibit the transition from quasiperiodic to chaotic dynamics. This has been corroborated by the maximal lyapunov exponent of the system. Furthermore, on investigating the correlation of the time-series using the Pearson's correlation coefficient, it is found that the chaotic system is anti-phase synchronized, whereas the quasiperiodic is not synchronized in any form. The parameteric regime where this transition has been observed is from the \textit{unbroken}-$\mathcal{PT}$ regime to the \textit{broken}-$\mathcal{PT}$ regime. Similarly, in the latter configuration with two amplified oscillators coupled to two attenuated oscillators, a similar transition has been observed. But in the neighbourhood of the \textit{Exceptional Point} ($\mathcal{EP}$) of the system, the system is shown to exhibit in-phase synchronized dynamics as is evident from the correlation analysis.
		\end{abstract}
			
		
	\end{frontmatter}
	
	
	\section{Introduction}
	\label{sec:sample1}
	Synchronization is perhaps one of the most prevalent drive in all of nature. It extends from
	the farthest reach of the cosmos to the sub-atomic scale. In the natural world, fireflies flashing
	in patterns to attract mates [1], rhythms in the pacemaker cells of our heart [2], hand clapping
	in a crowd [3], etc. are all events that display this force. In the technological realm, the atoms
	in a medium pulsating in synchrony is what gives rise to the coherent light known as laser [4].
	On the other hand, synchronization is not always a good thing. In a patient suffering from
	epilepsy, thousands of nerve cells in the brain discharge in a pathological manner leading to an
	epileptic attack [5]. Inanimate objects too could display this phenomenon. In fact, the London
	Millennium Bridge, traversing the River Thames, was closed two days after opening for safety
	reasons [6]. It was reported that the lateral motions caused by the pedestrians would cause the
	bridge to lurch to one side and as a result, the pedestrians would have to adjust their rhythms to keep themselves from falling down. After redesigning the bridges with additional dampers, the bridge was once again opened to the public. Hence, it could be seen that synchronization is a
	phenomenon that could be observed in almost all domains of this world.
	
	On the other hand, Carl M. Bender and his student Stefan Boettcher discovered that certain non-Hermitian Hamiltonians possess a real eigenspectra as long as they satisfy certain prerequisites [7-10]. These Hamiltonians came to be known as Parity and Time-reversal ($\mathcal{PT}$) Symmetric Hamiltonians and as such, they are invariant under the joint operation of the parity ($\mathcal{P}$) and time-reversal ($\mathcal{T}$) operator. The parity operator is a linear operator and it is defined as $\hat{x} \rightarrow -\hat{x}$ and $\hat{p} \rightarrow -\hat{p}$ and the time-reversal operator is an antilinear operator and it is defined as $\hat{x} \rightarrow -\hat{x}$, $\hat{p} \rightarrow -\hat{p}$ and $i \rightarrow -i$. One of the most interesting aspect of such Hamiltonians is the presence of $\mathcal{EP}$ in the eigenspectra. These are regions in the parameter space of the Hamiltonian where the real component of the eigenvalues coalesce thereby signifying a phase transition of the eigenspectra from real to imaginary. Their research marked a major milestone in the foundational studies of Quantum Mechanics. Not so long after the research group under the supervision of Prof. D. N. Christodoulides came up with the proposition that optics could provide the means for the experimental realization of the $\mathcal{PT}$-Symmetric quantum potentials [11]. And in 2012, R\"uter \textit{et al.} demonstrated the observation of $\mathcal{PT}$-Symmetry in a configuration of evanescently coupled waveguide structure with balanced gain and loss [12]. Since then, $\mathcal{PT}$-Symmetry has been investigated in complex optical potentials [13], optomechanics [14-15], optical lattices [16], microring lasers [17], solitons [18-20], wireless power transfer [21], multilayered structures [22-23], many-body ultracold systems [24], Li\'enard oscillators [25-27], Ikeda-type optical systems [28] and so on.
	
	In this article, we investigate the emergence of synchronization dynamics in two configurations of coupled $\mathcal{PT}$-symmetric nonlinear oscillators. Phenomena such as amplitude death [26] and extreme events [27] have been reported in such systems. Here, the two configurations consist of linearly coupled nonlinear oscillators with balanced amplification and attenuation so as to respect the conditions of $\mathcal{PT}$-symmetry. In section II, we discuss the theoretical modelling of such systems and the simulation results have been discussed in section III. This is followed by our conclusion in section IV.
	
	\section{Modelling and Results}
	
	\subsection{Two Oscillators Configuration}
	
	The equations governing the dynamics of the 2-oscillator system are as follows [25].
	
	\begin{subequations}
		\begin{align}
			& \frac{d^2x_1}{dt^2} - \gamma x^{2}_1 \frac{dx_1}{dt}+\beta x^3_1 + \alpha x_1+\kappa x_2 = 0\\
			& \frac{d^2x_2}{dt^2} + \gamma x^{2}_2 \frac{dx_2}{dt}+\beta x^{3}_2 +\alpha x_2+\kappa x_1 =0
		\end{align}
	\end{subequations}
	
	Here, $\gamma$ is the gain/loss coefficient, $\alpha = \omega_0^2$ is the natural frequency, $\beta$ is the coefficient of Duffing nonlinearity and $\kappa$ is the coupling constant of the two oscillators. All the parameters are positive real quantities. In such systems, the parity operator is defined as $x_1 \leftrightarrow x_2$ and the time reversal operator is defined as $t \rightarrow -t$. Under the simultaneous operator of both operators, the system, as a whole, remains invariant. Furthermore, it could be seen that one of the oscillators is amplified and the other is attenuated by the same proportion. Using $y_i=dx_i/dt$, this system could be transformed as follows. 
	
	\begin{subequations}
		\begin{align}
			& \frac{dx_1}{dt} = y_1\\
			& \frac{dy_1}{dt} = \gamma x^{2}_1 y_1 - \beta x^3_1 - \alpha x_1 - \kappa x_2\\
			& \frac{dx_2}{dt} = y_2\\
			& \frac{dy_2}{dt} = -\gamma x^{2}_2 y_2 - \beta x^3_2 - \alpha x_2 - \kappa x_1
		\end{align}
	\end{subequations}
	
	The fixed points of these equations are as follows.
	\begin{enumerate}
		\item \textbf{FP1} $\rightarrow$ $(x_1, y_1, x_2, y_2)=(0,0,0,0)$
		\item \textbf{FP2} $\rightarrow$ $(x_1, y_1, x_2, y_2)=(\pm a_1,0,\mp a_1,0)$
		\item \textbf{FP3} $\rightarrow$ $(x_1, y_1, x_2, y_2)=(\pm a_2,0,\pm a_2,0)$
	\end{enumerate}
	
	where $a_1=\sqrt{(\kappa-\alpha)/\beta}$ and $a_2 = \sqrt{(-\kappa-\alpha)/\beta}$. The stability of these fixed points could be ascertained by evaluating the eigenspectra of the Jacobian matrix given below.
	
	\begin{equation}
		J=
		\begin{pmatrix}  
			0 & 1 & 0 & 0 \\
			A & \gamma x^2_1 & -\kappa & 0 \\
			0 & 0 & 0 & 1 \\
			-\kappa & 0 & B & -\gamma x^2_2 \\
		\end{pmatrix}
	\end{equation}
	
	where $A=2 \gamma x_1 y_1 - \alpha - 3 \beta x^2_1$ and $B=-2 \gamma x_2 y_2 - \alpha - 3 \beta x^2_2$. For all the fixed point as mentioned above, it could be seen that $y_1=y_2=0$ and so, the Jacobian could be rewritten as follows.
	
	\begin{equation}
		J=
		\begin{pmatrix}  
			0 & 1 & 0 & 0 \\
			- \alpha - 3 \beta x^2_1 & \gamma x^2_1 & -\kappa & 0 \\
			0 & 0 & 0 & 1 \\
			-\kappa & 0 & - \alpha - 3 \beta x^2_2 & -\gamma x^2_2 \\
		\end{pmatrix}
	\end{equation}
	
	For our analysis of the eigenspectra of the Jacobian matrix, we have chosen $\gamma=0.1$, $\kappa=0.5$ and $\beta=1.0$. In our discussion, we will analyze the eigenspectra of the Jacobian for \textbf{FP1} and \textbf{FP2}. The reason for doing this is because for our choice of parameters, \textbf{FP3} will always be a purely imaginary quantity. On the other hand, \textbf{FP2} will also be an imaginary quantity for $\alpha > \kappa$. The trivial fixed point \textbf{FP1} is the simplest fixed point and the eigenvalues of the Jacobian for this fixed point are $\lambda_{1,2}=\pm\sqrt{0.5-\alpha}$ and $\lambda_{3,4}=\pm\sqrt{-0.5-\alpha}$. We are interested in the real component of the eigenvalues and it could be seen that the eigenvalues $\lambda_{3,4}$ are purely imaginary. But the eigenvalues $\lambda_{1,2}=\pm\sqrt{0.5-\alpha}$ are purely real for $\alpha<0.5$ and purely imaginary for $\alpha>0.5$. From these eigenvalues, it could be ascertained that at $\alpha=0.5$, the real component of all eigenvalues coalesce and thus, this could be termed as the $\mathcal{EP}$ of the system. Furthermore, it could be seen that the $\mathcal{EP}$ could be identified using the trivial fixed point of nonlinear oscillator systems. In Fig. 1, we have plotted the eigenvalues of the Jacobian for \textbf{FP2}. It could be seen that the real component of all four eigenvalues coalesce at $\alpha=0.5$, thereby signifying the credibility of our previous claim that we can evaluate the $\mathcal{EP}$ of the system by analyzing the regime where the eigenspectra of the Jacobian matrix depicts a transformation from real to imaginary..

	Below in Fig. 2, we have plotted the phase plane of the Gain Oscillator from $\alpha=0.35$ to $\alpha=0.6$. It could be seen that as α is increased, the chaotic attractor is seen to transform into a toroidal quasiperiodic attractor. And this is further validated by the bifurcation diagram of the temporal maxima $x_{1,max}$ and maximal Lyapunov exponent $\lambda_{max}$ in Fig. 3(a). As $\alpha$ is increased, $\lambda_{max}$ is seen to decrease in the neighbourhood of the $\mathcal{EP}$ to 0, thereby signifying the transition in the phase plane as observed in Fig. 2. So, from this, we can infer that our system exhibits the quasiperiodic route to chaos. 
	
	Our analysis of the time-series of the two oscillators reveals more interesting phenomenon. On plotting the time-series of the two oscillators in Fig. 3(b), it could be seen that the two oscillators are exhibiting distorted anti-phase synchronization of chaos. On evaluating the Pearson's Correlation Coefficient of the time-series of the oscillators for $\alpha=0.3$, the Correlation Coefficient is found to be $C=-0.973$ which validates our claim.
	
	\begin{subequations}
		\begin{align}
			C = \frac{\langle (x_{1,i} - \hat{x}_1) (x_{2,i} - \hat{x}_2)\rangle}{\sigma_1\sigma_2}
		\end{align}
	\end{subequations}
	
	And on analysing the Correlation Coefficient in the entire parametric regime of $\alpha$ in Fig. 5, it could be seen that $\alpha$ is increased, the two oscillators exhibit a phase transition from anti-phase synchronized chaotic dynamics to desynchronized quasiperiodic dynamics as is evident from Fig. 2. Furthermore, it must be noted here that in the $\mathcal{PT}$-Symmetric Optical Dimer, the spatial evolution of optical power in the two waveguides exhibit anti-phase synchronized periodic dynamics in the \textit{unbroken} $\mathcal{PT}$ Regime and in the \textit{broken} $\mathcal{PT}$ Regime, it exhibits exponential growth and decay of optical power leading to a total loss of synchronization [12]. So, in our $\mathcal{PT}$-Symmetric Nonlinear Oscillator system, we can say that this system exhibits $\mathcal{PT}$-Symmetry Breaking Induced Desynchronization of Temporal Dynamics.
	
	\subsection{4-Oscillators Configuration}
	
	We would now like to discuss the synchronization dynamics in a 4-oscillator configuration. A schematic of the configuration in shown in Fig. 6. The 4-oscillator configuration consists of two layers - one layer is composed of gain oscillators and the other is composed of loss oscillators. There is no coupling between the oscillators in one layer but the oscillators in different layers are coupled. Under such considerations, the mathematical model of the system could be given as follows.
	
	\begin{subequations}
		\begin{align}
			& \frac{d^2x_{1,1}}{dt^2} - \gamma x^{2}_{1,1} \frac{dx_{1,1}}{dt}+\beta x^3_{1,1} + \alpha x_{1,1}+\kappa (x_{1,2}+x_{2,2}) = 0 \\
			& \frac{d^2x_{2,1}}{dt^2} - \gamma x^{2}_{2,1} \frac{dx_{2,1}}{dt}+\beta x^3_{2,1} +\alpha x_{2,1}+\kappa (x_{1,2}+x_{2,2})= 0 \\
			& \frac{d^2x_{1,2}}{dt^2} + \gamma x^{2}_{1,2} \frac{dx_{1,2}}{dt}+\beta x^3_{1,2} + \alpha x_{1,2}+\kappa (x_{1,1}+x_{2,1}) = 0 \\
			& \frac{d^2x_{2,2}}{dt^2} + \gamma x^{2}_{2,2} \frac{dx_{2,2}}{dt}+\beta x^3_{2,2} +\alpha x_{2,2}+\kappa (x_{1,1}+x_{2,1})= 0
		\end{align}
	\end{subequations}
	
	If we consider the absence of any form of nonlinearity in the system, then this mathematical model could be rewritten as follows.
	
	\begin{subequations}
		\begin{align}
			& \frac{d^2x_{1,1}}{dt^2} + \alpha x_{1,1}+\kappa (x_{1,2}+x_{2,2}) = 0 \\
			& \frac{d^2x_{2,1}}{dt^2} + \alpha x_{2,1}+\kappa (x_{1,2}+x_{2,2})= 0 \\
			& \frac{d^2x_{1,2}}{dt^2} + \alpha x_{1,2}+\kappa (x_{1,1}+x_{2,1}) = 0 \\
			& \frac{d^2x_{2,2}}{dt^2} + \alpha x_{2,2}+\kappa (x_{1,1}+x_{2,1})= 0
		\end{align}
	\end{subequations}
	
	And using $dx_{i,j}/dt=y_{i,j}$, we can further simplify this system of equations.
	
	\begin{equation}
		\frac{d}{dt} \begin{pmatrix}
			x_{1,1} \\ y_{1,1} \\ x_{2,1} \\ y_{2,1} \\ x_{1,2} \\ y_{1,2} \\ x_{2,2} \\ y_{2,2}
		\end{pmatrix} = M \begin{pmatrix}
			x_{1,1} \\ y_{1,1} \\ x_{2,1} \\ y_{2,1} \\ x_{1,2} \\ y_{1,2} \\ x_{2,2} \\ y_{2,2}
		\end{pmatrix}
	\end{equation}
	
	where the matrix $M$ is given by
	
	\begin{equation}
		M = \begin{bmatrix}
			0 & 1 & 0 & 0 & 0 & 0 & 0 & 0\\
			-\alpha & 0 & 0 & 0 & -\kappa & 0 & -\kappa & 0\\
			0 & 0 & 0 & 1 & 0 & 0 & 0 & 0\\
			0 & 0 & -\alpha & 0 & -\kappa & 0 & -\kappa & 0\\
			0 & 0 & 0 & 0 & 0 & 1 & 0 & 0\\
			-\kappa & 0 & -\kappa & 0 & -\alpha & 0 & 0 & 0\\
			0 & 0 & 0 & 0 & 0 & 0 & 0 & 1\\
			-\kappa & 0 & -\kappa & 0 & 0 & 0 & -\alpha & 0
		\end{bmatrix}
	\end{equation}

	The eigenvalues of this matrix for $\kappa=0.1$ are $\lambda_1=\lambda_2=\sqrt{-\alpha}$, $\lambda_3=\lambda_4=-\sqrt{-\alpha}$, $\lambda_{5, 6}=\pm\sqrt{-\alpha-1/5}$ and $\lambda_{7, 8}=\sqrt{-\alpha+1/5}$. We can see that the eigenvalues $\lambda_1$ to $\lambda_6$ are all purely imaginary quantities. But the eigenvalues $\lambda_7$ and $\lambda_8$ will become purely imaginary when $\alpha > 0.2$. From this, we can conclude that $\alpha = 0.2$ is the $\mathcal{EP}$ of the system as is done previously. A plot of the real component of all the eigenvalues shall depict that at $\alpha = 0.2$, there is coalescence of the eigenvalues of the matrix $M$. So, we will now analyze the temporal dynamics of one of the Gain oscillators in this parametric regime.

	In Fig. 7, the temporal evolution of the gain oscillator $x_{1,1}$ for different values of the natural frequency of the oscillator $\alpha$ is shown. For $\alpha=0.05$, it can be seen to be in a chaotic state, which eventually transforms to quasiperiodic state as $\alpha$ is increased further. This has been further corroborated in the phase plane of the oscillator in Fig. 8. Similar behavior was also seen in the 2-oscillator configuration. Furthermore, on analyzing the synchronization dynamics of the oscillators in the same layer in Fig. 9, it can been seen that the oscillators in the same layer are in-phase synchronized in the absence of any form of coupling between them. But anti-phase synchronization is observed between oscillators in different layers when $\alpha=0.05$ but this behavior disappears for $\alpha=0.25$. To further elucidate the chaotic time-series of the gain oscillator $x_{1,1}$, we have plotted the $\lambda_{max}$ of the time-series for the gain oscillator in Fig. 10. It could seen that the $\lambda_{max}$ abruptly decreases to zero at $\alpha=0.16$ thereby signifying the transition in temporal dynamics from chaotic to quasiperiodic as is observed in the time-series in Fig. 7.
	
	Now, to analyze the synchronization dynamics of the oscillators $x_{1,1}$ and $x_{1,2}$, we have plotted the Pearson's Correlation Coefficient in Fig. 11. It could be seen for $\alpha<0.2$, the correlation cofficient is close to $-1.0$ signifying the anti-phase synchronized chaotic dynamics. But as we increase $\alpha$, it starts increasing and in the neighbourhood of the $\mathcal{EP}$, it shoots up to 1.0 which implies that the oscillators in both layers are in-phase synchronized. But as we increase it beyond the $\mathcal{EP}$, it abruptly decreases to zero which implies that synchronization between the oscillators in the two layers is evidently lost.
	
	So, from our analysis, one important conclusion that can be drawn is that beyond the $\mathcal{EP}$ in both 2-oscillator and 4-oscillator configurations, synchronization dynamics between the oscillators disappears beyond the $\mathcal{EP}$ of the system. In $\mathcal{PT}$-Symmetry, this regime is defined as the \textit{broken}-$\mathcal{PT}$ regime and as such, from our analysis, we can claim that our system of coupled nonlinear oscillators exhibit the $\mathcal{PT}$-Symmetry breaking induced loss of synchronization in coupled nonlinear oscillator systems.
	
	\section{Conclusion}
	We discussed the dynamics in the synchronization of $\mathcal{PT}$-Symmetric nonlinear oscillators for two configurations. In the former, two oscillators are coupled with each other. In the latter, we have two layers of oscillators - one comprises the gain oscillators and the other comprixes the lossy oscillators. On analyzing the real component of the eigenspectra of the linearization Jacobian in the first configuration, we discovered the presence of $\mathcal{EP}$ in the eigenspectra and as such, we defined the \textit{unbroken} and \textit{broken} $\mathcal{PT}$-regime. On analyzing the temporal evolution of one of oscillator in both regimes and we discovered that the oscillators exhibit anti-phase synchronization of chaotic dynamics in the \textit{unbroken} regime, whereas in the \textit{broken} regime, the oscillators loses synchronized dynamics and there is a phase transition in the temporal dynamics from chaotic to quasiperiodic. This has been further corroborated by the maximal Lyapunov exponent and the Pearson's correlation coefficient of the time-series data of the system. In the 2nd configuration, the oscillators in the same layers exhibit in-phase chaotic and quasiperiodic dynamics in both regimes. Oscillators in different layers exhibit anti-phase synchronized chaotic and desynchronized quasiperiodic dynamics. But in the neighbourhood of the $\mathcal{EP}$, we observed the emergence of in-phase synchronized dynamics of the oscillators in different layers. In a nutshell, the phase transition in the temporal dynamics from anti-phase synchronized chaotic to desynchronized quasiperiodic dynamics in the parameter space from \textit{unbroken} $\mathcal{PT}$-regime to the \textit{broken} $\mathcal{PT}$-regime leads us to conclude that this system exhibits the $\mathcal{PT}$-Symmetry Breaking induced loss of synchronization.
	
	
	\bibliographystyle{elsarticle-num} 
	\bibliography{cas-refs}
	\begin{enumerate} [label={[\arabic*]}]
		\item J. Buck and E. Buck. Mechanism of rhythmic synchronous flashing of fireflies. Fireflies of Southeast Asia may use anticipatory time-measuring in synchronizing their flashing, Science 159, 1319 (1968).
		\item D.C. Michaels, E.P. Matyas, and J. Jalife. Mechanisms of sinoatrial pacemaker synchronization: a new hypothesis, Circulation Res. 61, 704, (1987).
		\item Z. Néda, E. Ravasz, Y. Brechet, T. Vicsek, and A. -L. Barabási. The sound of many hands clapping, Nature 403, 849 (2000).
		\item R. Roy and K. S. Thornburg Jr.. Experimental synchronization of chaotic lasers, Phys. Rev. Lett. 72, 2009 (1994).
		\item P. Jiruska et al. Synchronization and desynchronization in epilepsy: controversies and hypotheses, J. Physiol. 591, 787 (2013).
		\item S. H. Strogatz et al. Crowd synchrony on the Millennium Bridge, Nature 438, 43 (2005).
		\item C. M. Bender, S. Boettcher. Real Spectra in Non-Hermitian Hamiltonians Having 
		PT-Symmetry, Phys. Rev. Lett. 80, 5243 (1998).
		\item C. M. Bender, S. Boettcher, P. N. Meisinger. PT-symmetric quantum mechanics, J. Math. Phys. 40, 2201 (1999).
		\item C. M. Bender, D.C. Brody, H. F. Jones, Phys. Rev. Lett. 89, 270401 (2002).
		\item C. M. Bender. Making sense of non-Hermitian Hamiltonians, Rep. Prog. Phys \textbf{70}, 947 (2007).
		\item R. El-Ganainy, K. G. Makris, D. N. Christodoulides and Z. H. Musslimani. Theory of coupled optical PT-symmetric structures, Opt. Lett. 32, 2632 (2007).
		\item C. E. R\"uter \textit{et al.}. Observation of parity–time symmetry in optics, Nat. Phys. 6, 192 (2010).
		\item Z. Lin \textit{et al.} Unidirectional Invisibility Induced by PT-Symmetric Periodic Structures, Phys. Rev. Lett. 106, 213901 (2011).
		\item X. Xu, Y. Liu, C. Sun, and Y. Li. Mechanical PT-symmetry in coupled optomechanical systems, Phys. Rev. A 92, 013852 (2015).
		\item X. L\"u, H. Jing, J. Ma, and Y. Wu. PT-Symmetry-Breaking Chaos in Optomechanics, Phys. Rev. Lett. 114, 253601 (2015).
		\item M. -A. Miri, A. Regensburger, U. Peschel and D. N. Christodoulides, Phys. Rev. A \textbf{86}, 023807 (2012).
		\item J. Ren \textit{et al.} Unidirectional light emission in PT-symmetric microring lasers, Opt. Express 26, 27153 (2018).
		\item A. K. Sarma, M.-A. Miri, Z. H. Musslimani, D. N. Christodoulides. Continuous and discrete Schrödinger systems with parity-time-symmetric nonlinearities, Phys. Rev. E 89, 052918 (2014).
		\item M. -A. Miri \textit{et al.} Bragg solitons in nonlinear PT-symmetric periodic potentials, Phys. Rev. A 86, 033801 (2012).
		\item A. Govindarajan, A. K. Sarma and M. Lakshmanan. Tailoring PT-symmetric soliton switch, Opt. Lett. 44, 663  (2019).	
		\item S. Assawaworrarit, X. Yu and S. Fan. Robust wireless power transfer using a nonlinear parity–time-symmetric circuit, Nat. 546, 387 (2017).
		\item M. Sarisaman. Unidirectional reflectionlessness and invisibility in the TE and TM modes of a PT-symmetric slab system, Phys. Rev. A 95, 013806 (2017).
		\item J. P. Deka and A. K. Sarma. Highly Amplified Light Transmission in Parity-Time Symmetric Multilayered Structure, Appl. Opt. 57, 1119 (2018).
		\item Y. Takasu \textit{et al.}. PT-symmetric non-Hermitian quantum many-body system using ultracold atoms in an optical lattice with controlled dissipation, Progress of Theoretical and Experimental Physics 2020, 12 (2020).
		\item J. P. Deka, A. K. Sarma, A. Govindarajan, M. Kulkarni. Multifaceted nonlinear dynamics in PT-symmetric coupled Liénard oscillators, Nonlinear Dyn 100, 1629 (2020).	
		\item U. Singh, A. Raina, V. K. Chandrasekar, and D. V. Senthilkumar. Nontrivial amplitude death in coupled parity-time-symmetric Liénard oscillators, Phys. Rev. E 104, 054204 (2021).
		\item B. Thangavel, S. Srinivasan and T. Kathamuthu. Extreme events in a forced BVP oscillator: Experimental and numerical studies, Chaos, Solitons and Fractals 153, 1 (2021).
		\item J. P. Deka, A. K. Sarma. Chaotic dynamics and optical power saturation in parity–time (PT) symmetric double-ring resonator, Nonlinear Dyn 96, 565 (2019).
	\end{enumerate}

\begin{figure}
	\centering
	\includegraphics[height=7cm,width=8cm]{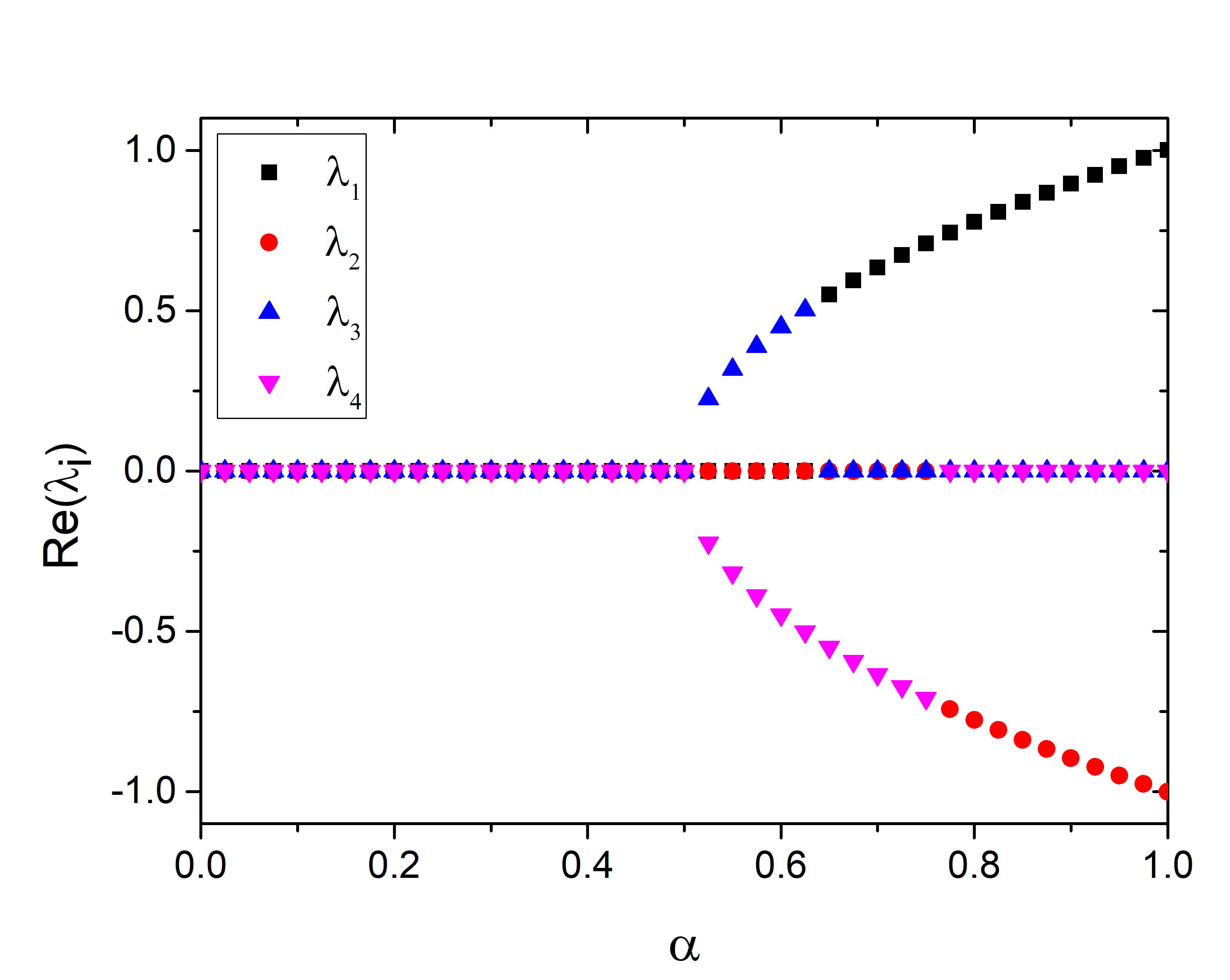}
	\caption{Eigenspectra of the Jacobian Matrix for fixed point \textbf{FP2} as a function of $\alpha$. Other parameters: $\gamma=0.1$, $\kappa=0.5$ and $\beta=1.0$.}
\end{figure}

\begin{figure}
	\includegraphics[height=10cm,width=16cm]{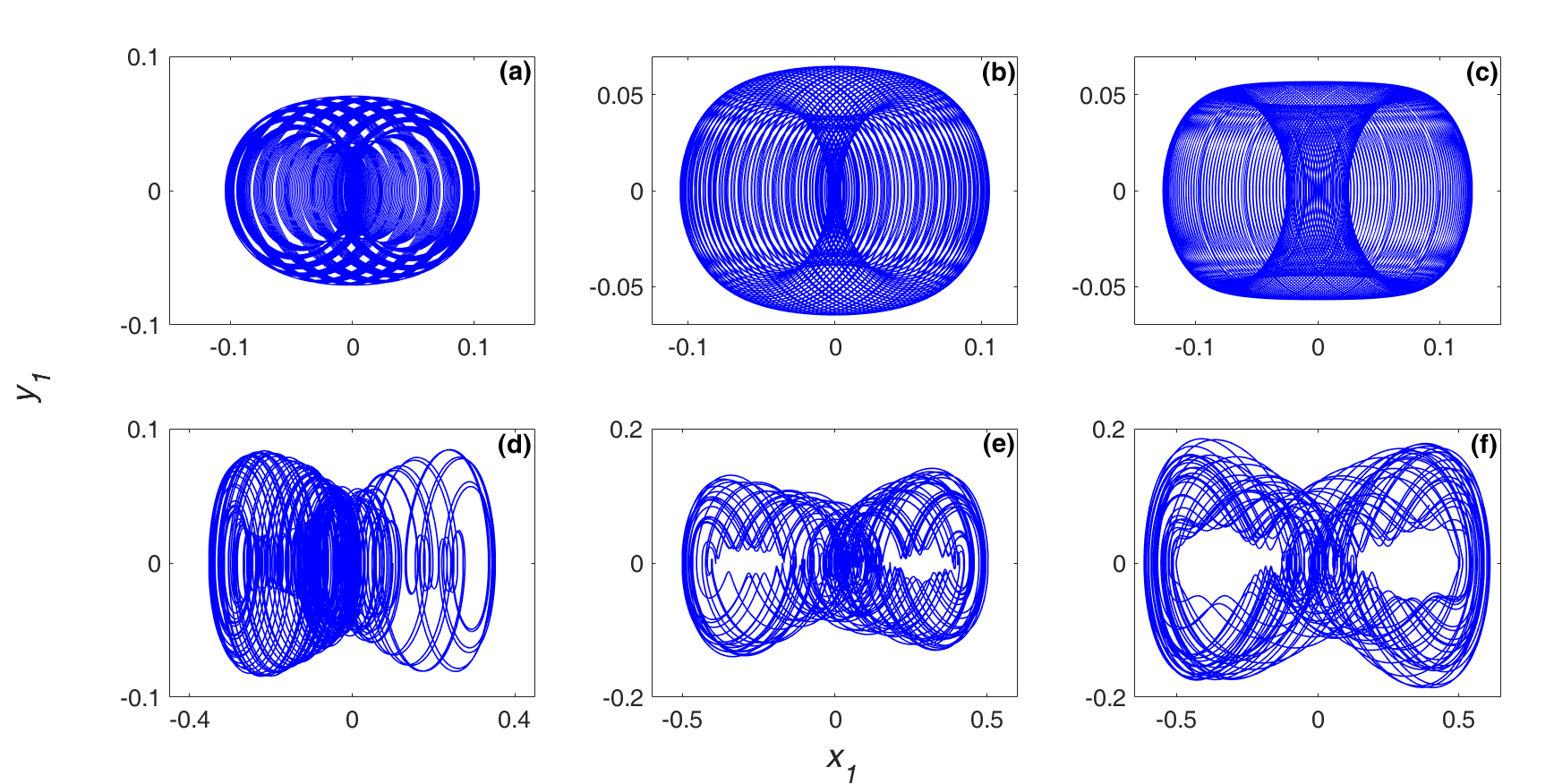}
	\caption{Phase Plane of the Gain Oscillator of the 2-oscillator configuration for (a) $\alpha=0.6$, (b) $\alpha=0.55$, (c) $\alpha=0.5$, (d) $\alpha=0.45$, (e) $\alpha=0.4$ and (f) $\alpha=0.35$. Other Parameters - $\gamma=0.1$, $\kappa=0.5$ and $\beta=1$.}
\end{figure}

\begin{figure}
	\centering
	\includegraphics[height=7cm,width=15cm]{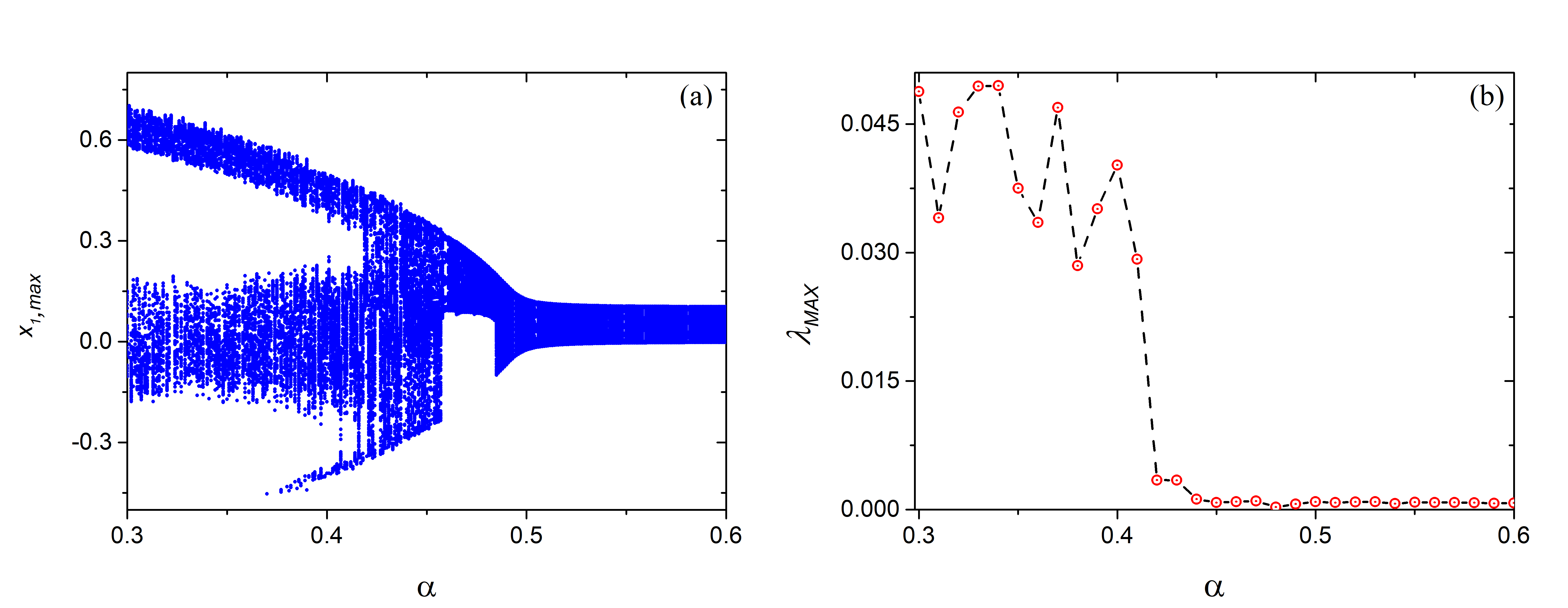}
	\caption{(a) Bifurcation Diagram and (b) Maximal Lyapunov Exponent v/s. the natural frequency of the 2-oscillator configuration for $\gamma=0.1$, $\kappa=0.5$ and $\beta=1.0$.}
\end{figure}

\begin{figure}
	\centering
	\includegraphics[height=7cm,width=9cm]{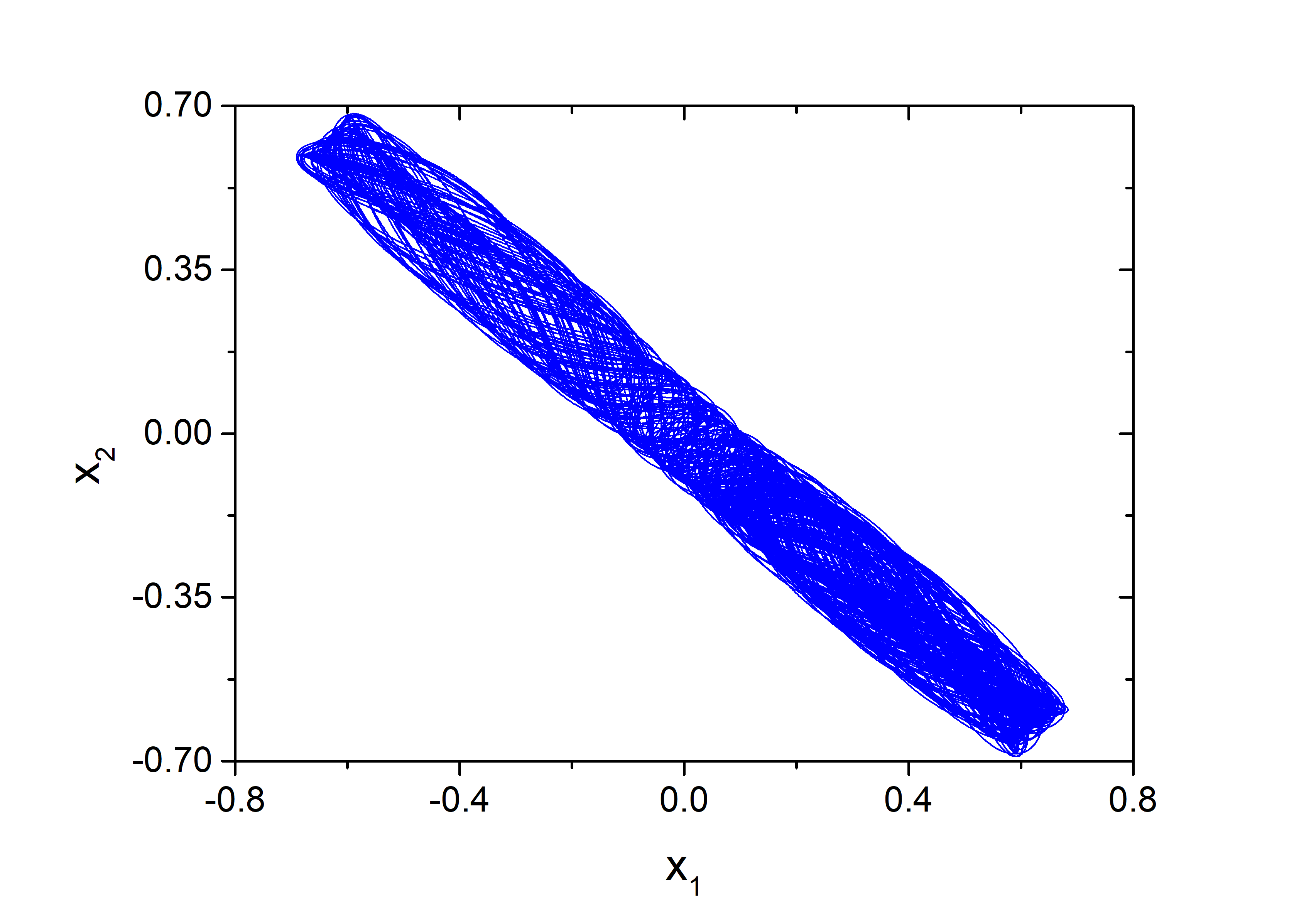}
	\caption{Temporal Evolution of the Gain Oscillator vs. Temporal Evolution of the Lossy Oscillator for $\alpha=0.3$, $\gamma=0.1$, $\kappa=0.5$ and $\beta=1.0$.}
\end{figure}

\begin{figure}
	\centering
	\includegraphics[height=7cm,width=9cm]{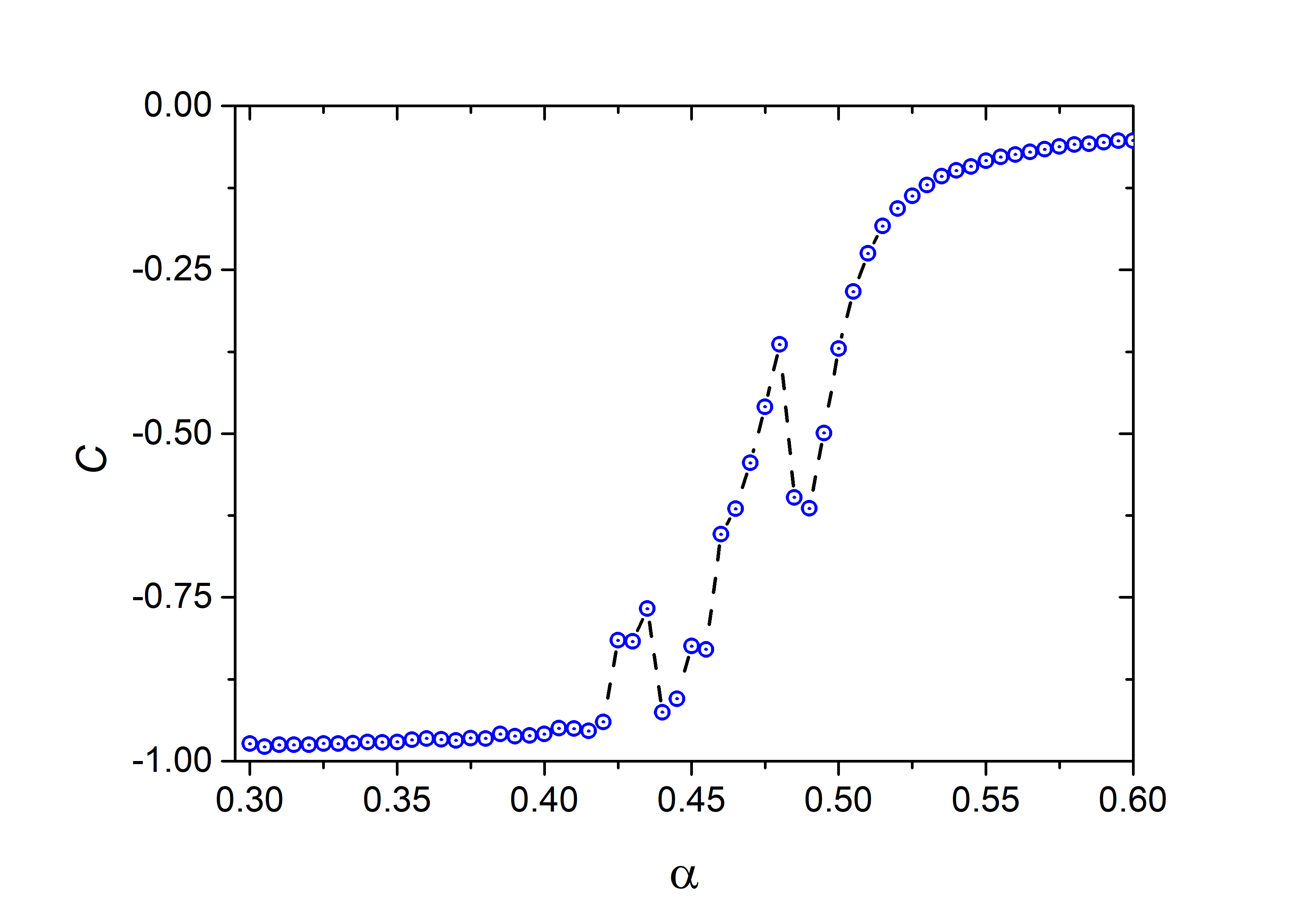}
	\caption{Pearson's Correlation Coefficient $C$ vs. Natural Frequency of the Oscillators α for $\gamma=0.1$, $\kappa=0.5$ and $\beta=1.0$.}
\end{figure}

\begin{figure}
	\centering
	\includegraphics[height=7cm,width=9cm]{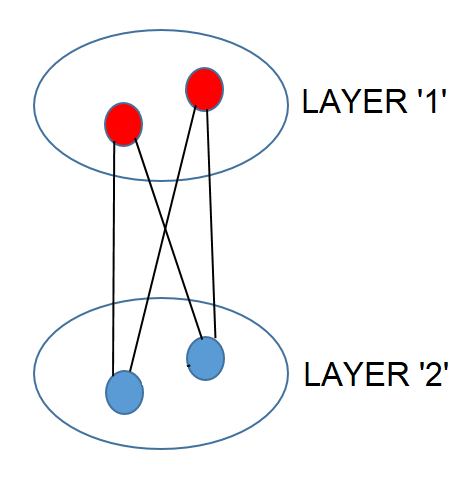}
	\caption{Schematic of the 4-oscillators configuration.}
\end{figure}

\begin{figure}
	\centering
	\includegraphics[height=10cm,width=16cm]{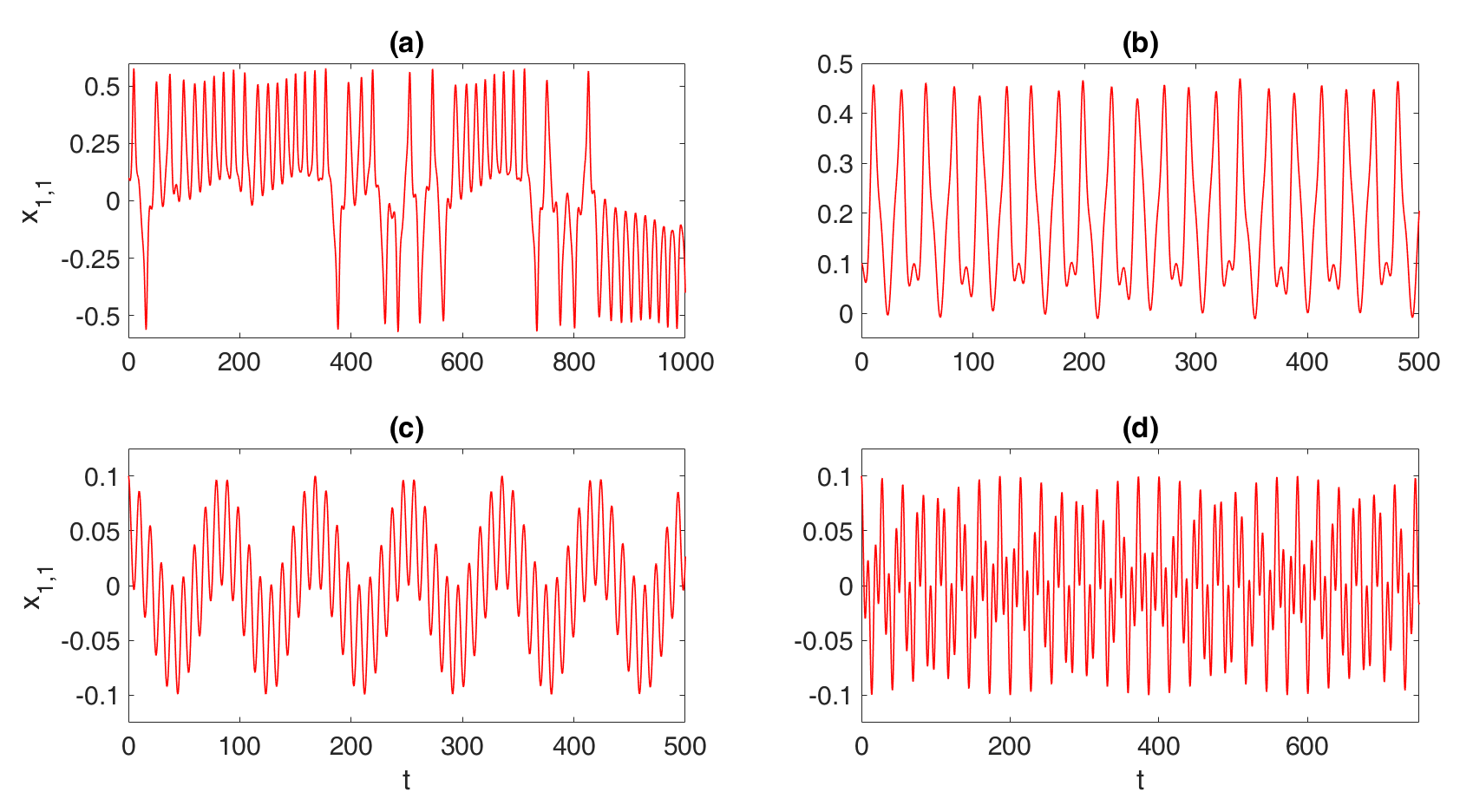}
	\caption{Temporal Evolution of the Gain Oscillator $x_{1,1}$ for (a) $\alpha=0.05$, (b) $\alpha=0.1$, (c) $\alpha=0.2$ and (d) $\alpha=0.25$.}
\end{figure}

\begin{figure}
	\centering
	\includegraphics[height=6cm,width=16cm]{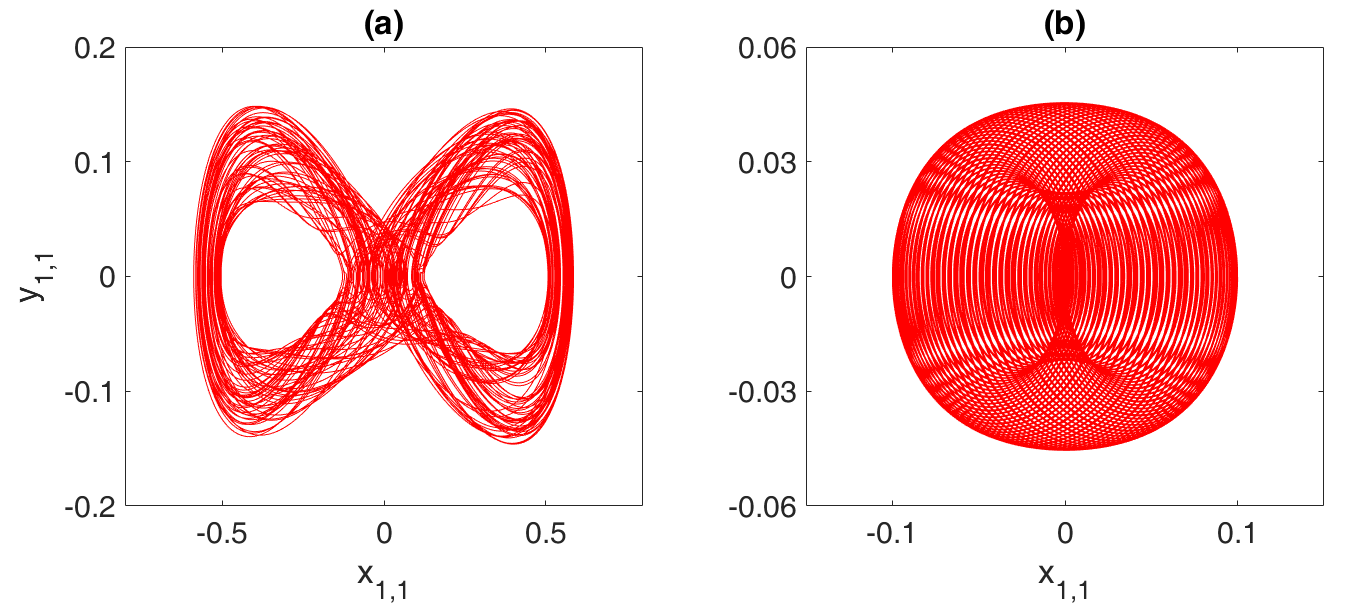}
	\caption{Phase Plane of the Gain Oscillator of the 4-oscillator configuration $x_{1,1}$ for (a) $\alpha=0.05$ and (b) $\alpha=0.25$.}
\end{figure}

\begin{figure}
	\centering
	\includegraphics[height=8cm,width=8cm]{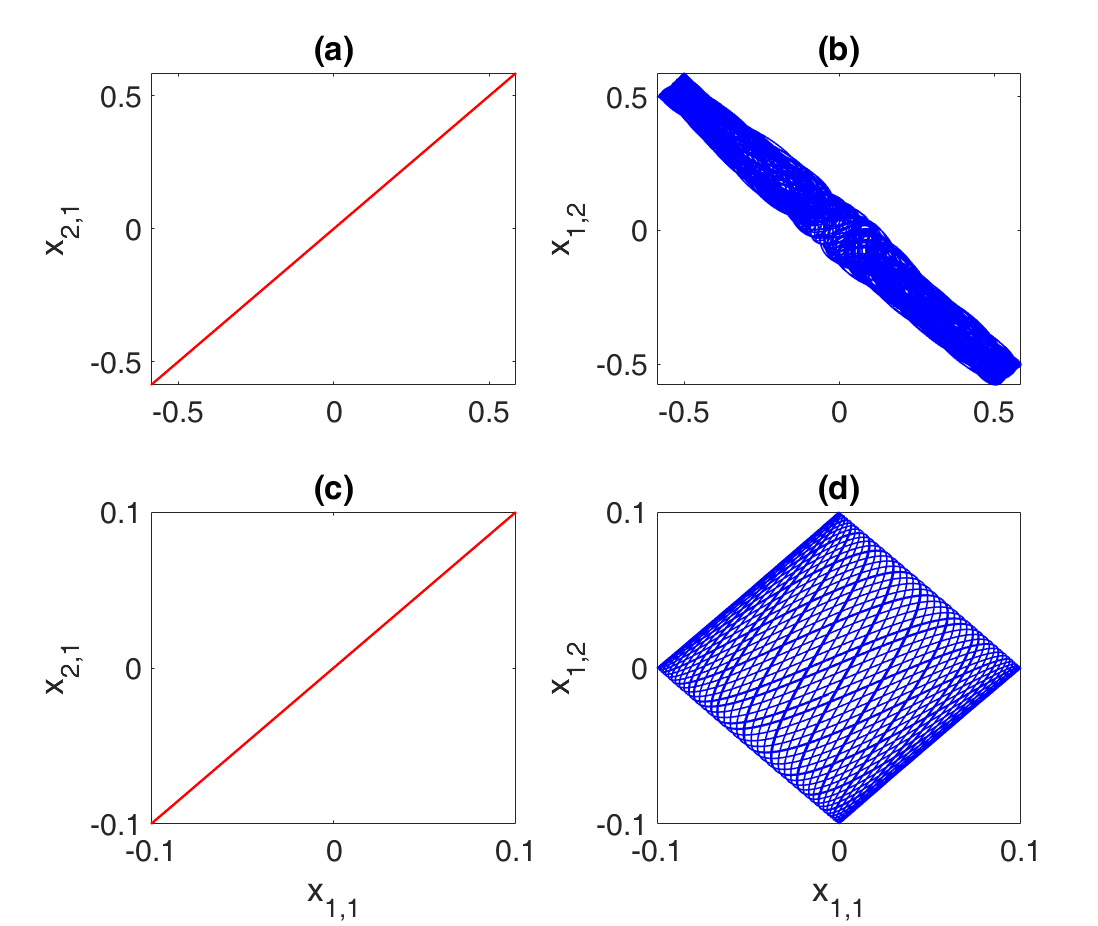}
	\caption{Temporal Evolution of the Gain Oscillators $x_{1,1}$ and $x_{1,2}$ and	Temporal of the Gain Oscillator $x_{1,1}$ and Lossy Oscillator $x_{1,2}$ for (a-b) $\alpha=0.05$ and (c-d) $\alpha=0.25$.}
\end{figure}

\begin{figure}
	\centering
	\includegraphics[height=6cm,width=8cm]{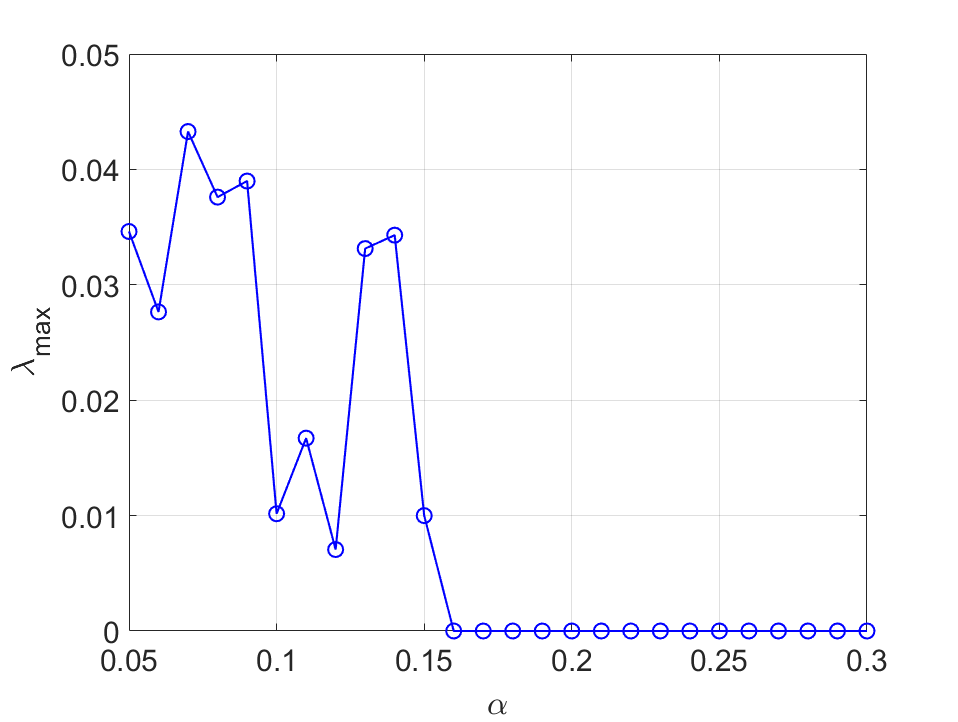}
	\caption{Maximal Lyapunov Exponent v/s. the natural frequency of the time-series of the gain oscillator $x_{1,1}$ for $\gamma=0.1$, $\kappa=0.1$ and $\beta=1$.}
\end{figure}

\begin{figure}
	\centering
	\includegraphics[height=7cm,width=8cm]{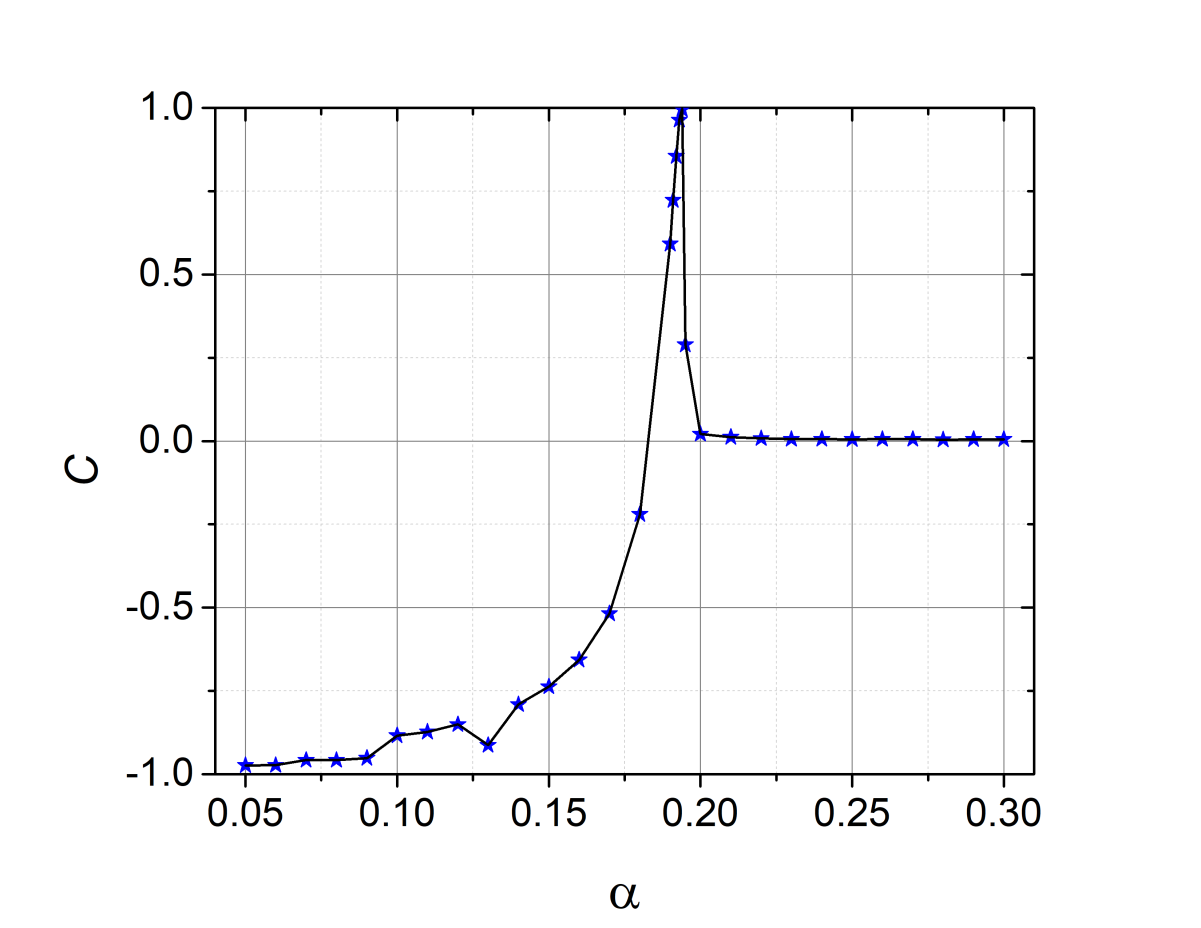}
	\caption{Pearson's Correlation Coefficient $C$ of the temporal evolution of $x_{1,1}$ and $x_{1,2}$ v/s $\alpha$.}
\end{figure}

\end{document}